\documentstyle[prl,aps,multicol,epsf]{revtex}

\newcommand{\half}{\frac{1}{2}}
\newcommand{\E}{\varepsilon}
\newcommand{\sts}{
   \parbox{1mm}{
      \setlength{\unitlength}{0.7mm}
      \begin{picture}(1,9)
         \thinlines
         \put(0.5,0){\line(0,1){7.7}}
      \end{picture}
   }
}

\newcommand{\lts}{
   \parbox{1mm}{
      \setlength{\unitlength}{0.9mm}
      \begin{picture}(1,12)
         \thinlines
         \put(0.5,-0.2){\line(0,1){11}}
      \end{picture}
   }
}
\newcommand{\R}{\mbox{\rm I}\!\mbox{\rm R}}

\newcommand{\rme}{\mbox{e}}

\newcommand{\be}{\begin{equation}}
\newcommand{\ee}{\end{equation}}
\newcommand{\bea}{\begin{eqnarray}}
\newcommand{\eea}{\end{eqnarray}}
\newcommand{\nn}{\nonumber}

\newcommand{\nicht}[1]{}
\newcommand{\eq}[1]{(\ref{#1})}

\newlength{\pcm}
\newlength{\pmm}
\newcommand {\GN} {\,\epsfxsize=0.13\pcm \parbox{0.13\pcm}{\epsfbox{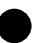}}\,}

\newcommand{\DynG}{{\,\epsfxsize=1.28\pcm \parbox{1.28\pcm}{\epsfbox{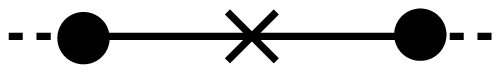}}\,}}
\newcommand{\DynH}{{\,\epsfxsize=1.44\pcm \parbox{1.44\pcm}{\epsfbox{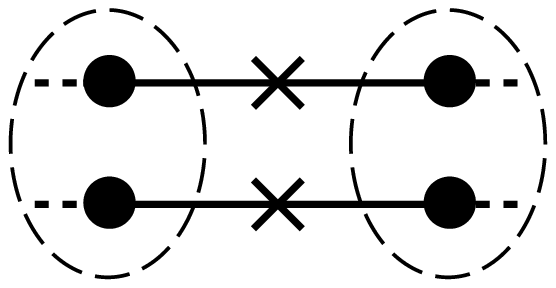}}\,}}
\newcommand{\DynI}{{\,\epsfxsize=0.8\pcm \parbox{0.8\pcm}{\epsfbox{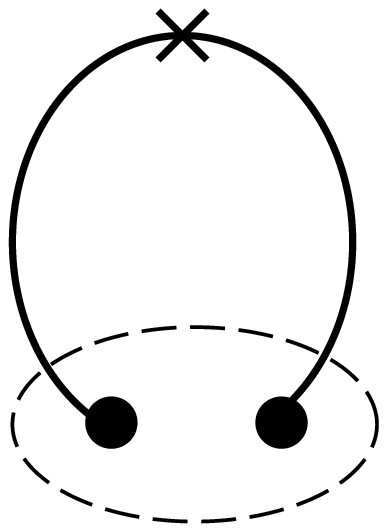}}\,}}

\title{Dynamics of Selfavoiding Tethered Membranes II\\
Inclusion of Hydrodynamic Interaction (Zimm Model)}
\author{\centerline{Kay J\"org Wiese}
\newline
\centerline{Fachbereich Physik, Universit\"at GH Essen,  45117 Essen,
Germany}}
\date{\today}
\begin{document}
\maketitle

\begin{abstract}
The dynamical scaling properties of selfavoiding polymerized membranes
with internal dimension $D$ embedded into $d$ dimensions
are studied including hydrodynamical interactions. 
It is shown that the theory is renormalizable to all orders in 
perturbation theory and that the dynamical scaling exponent
$z$ is given by $z=d$. The crossover to the region, where the
membrane is crumpled swollen but the hydrodynamic interaction 
irrelevant is discussed. The results  apply as well  
to polymers ($D=1$) as to membranes ($D=2$).
\end{abstract}

\begin{multicols}{2}
\narrowtext
The statistical properties of polymerized flexible membranes, 
generalizing polymers, have found large interest during the
last years. Due to the selfavoidance, they are either
found in a flat or crumpled swollen phase \cite{AbrEtAl89,AbrNel90,GrestMurat90,GrestPetsche94,KrollGompper93,HwaKokufutaTanaka,SpectorEtAl94}. 
An analytical approach was initiated in \cite{AroLub87,KarNel87}, 
where calculations of the static scaling exponent $\nu$ 
were performed at 1-loop order.
Its consistency to all orders in perturbation theory has been established in \cite{DDG3,DDG4}.
Recently, 2-loop calculations have been performed, which give 
reliable results for all imbedding dimensions \cite{DavidWiese96a,WieseDavid96b}.

In this letter we want to adress the question of the dynamics of
such membranes in the crumpled swollen phase 
embedded in some (viscous) solvent.
(For a discussion of the flat phase, see \cite{FreyNelson91}.)

We first summarize the main results before discussing the 
technical procedure to derive them. 

First of all, the Brownian motion of the particles,
both of the solvent and of the fluid, have to be taken
into account.  It is responsible for the relaxation of 
	the membrane. This can be studied via the auto-correlation
function which has for large membrane size and large time
the scaling form
\be
\left< (r(x,t)-r(x,0))^2 \right> \sim t^{2/z}
\ee
Our goal is to determine $z$. 

In the physical system, hydrodynamics may be important.
Two cases can be distinguished: In the first case, the 
hydrodynamic is irrelvant and the exponent $z$ is given 
by
\be
z=2+d_f \ ,
\ee
where $d_f$ is the fractal dimension of the membrane. 
This result has been established to all orders in perturbation
theory in \cite{Wiese97a}. 
In the second case, hydrodynamic interactions are relevant and 
we will show below that this  
modifies the exponent $z$ to
\be
 z=d \ ,
\ee
where $d$ is the dimension of the embedding space, see below.
 This situation is
plotted in figure 1, where also the phase-separation line 
is given.

\begin{figure}[t] \label{f:phase-diagram}
\centerline{
\epsfxsize=8.7cm \parbox{8.7cm}{\epsfbox{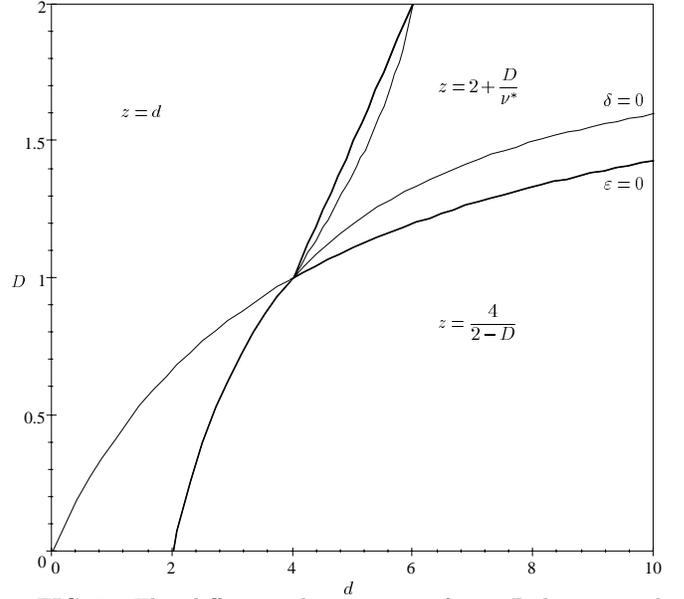}}
}
\caption{The different phase-regions for a $D$-dimensional membrane
embedded into $d$ dimensions including hydrodynamique interactions.
  The region with $\delta<0$ and $\E<0$ is the Gaussian phase.
Selfavoidance and hydrodynamic interaction are irrelevant, i.e.\ 
$\nu^*=\frac {2-D}2$ and $z=\frac 4 {2-D}$. Hydrodynamics is relevant
for small $d$ and becomes irrelevant if 
$d> d_f+2$, where $d_f=D/\nu^*$ is the fractal dimension of the membrane. 
For $d>d_f+2$ and $\E>0$, $z=2+d_f$.
}
\end{figure}%
The exponent $z$ should be observable via dynamic light-
or neutron-scattering methods. To our knowledge, no such 
experiment has been performed. 

Theoretically, the dynamics for polymers has first been regarded in \cite{DeGennes75,DeGennes75b}
using scaling arguments. For membranes, a similar analysis has been 
performed in \cite{KantorKardarNelson1986b}.%

Scaling can best be studied using renormalization group methods.
For polymers, such a treatment  
has been carried out  at 1-loop order in 
\cite{DeGennes75,DeGennes75b,AlNoaimiEtAl78,OonoFreed81,Oono85,PuriSchaubOono86,WangFreed86,SchaubCreamerJo1988}. A proof of the renormalizability
which ensures the correctnes of the method has been given in \cite{Wiese97a}
for the case of purley dissipative motion (Rouse model).

Let us now introduce the model with hydrodynamic interactions,
show that it is  renormalizable 
and calculate the scaling exponent $z$.

The static behavior of the membrane is given by the 
Edwards Hamiltonian
\begin{equation} \label{e:Ham}
\!\!\!\!{\cal H}[ r]= \frac{1}{2}\int_x\!\Big(\nabla  r(x)\Big)^2
+ b  \int_x\!\int_y\!
\delta^d\Big ( r(x)- r(y)\Big ) \ .
\end{equation}
The embedding of the $D$-dimensional membrane in $d$-dimensional bulk space
is described by the mapping $x\in \R^D\to  r(x)\in \R^d$.
$b$ is the coupling constant, associated to the self-avoidance.


Hydrodynamic interactions for polymers were first introduced
 by Zimm \cite{Zimm56}. He wrote down the following 
Langevin equation, which we will also use for membranes:
\be \label{LE}
	\dot r(x,t) = {\cal D} \cdot \left( -\frac{\delta H}{\delta r} +\zeta \right) (x,t)
\ee
Here, $\cdot$ denotes the scalar product of the 
matrix operator $\cal D$ and the vector $\delta H/\delta r$,
which is defined by
\be
f \cdot g := \int_x f_\alpha(x) g_\alpha(x) \ .
\ee 
The hydrodynamic interaction is
\bea \label{hydro}
&&{\cal D}_{\alpha \beta}(x,y,r(x,t),r(y,t)) = \\
&&\quad \lambda \delta_{\alpha\beta}\delta^D(x-y) + \lambda\eta\int_k \left(\frac{\delta_{\alpha\beta}}{k^2}-
\frac{k_\alpha k_\beta}{k^4}\right)
\rme^{ik(r(x,t)-r(y,t))} \nn
\eea
We will not repeat the  derivation \cite{Zimm56} of equation \eq{hydro} here. 
Let us however note that one 
 supposes that the hydrodynamic degrees of freedom are 
fast enough, so that their dynmaics can be neglected and that 
screening effects are irrelvant. This might be wrong for 
 membranes and in this case our results would only 
apply to membranes with large holes. 
For $\eta=0$, \eq{LE} reduces to purely diffusive motion (Rouse model).

The noise  correlation is 
\be
\overline{\zeta_\alpha(x,t)  ({\cal D}\cdot
\zeta)_\beta(y,t')} = 2\delta(t-t')  \delta^D(x-y) \ .
\ee
This ensures that the static behavior is correctly reproduced. 
Following Janssen \cite{Janssen92}, the corresponding field-theory 
is obtained by imposing the Langevin-equation through  an 
auxiliary field $\tilde r$.  Integration over the noise then
yields the dynamic functional in Ito-discretization
\bea \label{J}
J &=& \int_t \tilde r \cdot \dot r + \tilde r \cdot {\cal D} \cdot \frac{\delta H}{\delta r} - \tilde r \cdot {\cal D} \cdot \tilde r \ .
\eea
This model has to be renormalized. Analogously to \cite{Wiese97a} 
divergences only occur at small distances. They can be analyzed via 
a multilocal operator product expansion (MOPE). Renormalizability is ensured
\cite{Wiese97a,DDG3,DDG4} if counter-terms for all 
possible marginal and relevant operators are included into the
action. Due to causality, only operators with at least one 
response field $\tilde r$ are needed. These are the local 
operators $\tilde r \dot r$, $\tilde r (-\Delta) r$ and $\tilde r^2$.
Other local operators like $\tilde r r^n$ are forbidden by translation-invariance in $r$-space.

By the same arguments one finds that there are no new marginal or relevant
counter-terms proportional to 
 2- or 3- body interactions and 4-body-interactions
are irrelvant. (Of  course, long-range interactions are relevant, but
they are not generated in perturbation theory.)
We now want to show that the structure of the model 
is preserved, i.e.\ that it can be renormalized if we introduce
renormalized quantities according to
\bea
\lambda &=& Z_\lambda \lambda_R \sqrt{\frac Z {\tilde Z}} \nn\\
r &=& \sqrt Z r_R \nn\\
\tilde r &=& \sqrt{\tilde Z} \tilde r_R \\
b &=& b_R Z_b Z^{d/2} \mu^{\E} \nn \\
\eta &=& \eta_R Z_\eta Z_\lambda^{-1} Z^{d/2-1} \mu^{\delta} \nn 
\eea
The two regularization parameters $\E$ and $\delta$ are given 
by
\bea
\E &=& 2D - \nu d \nn \\ 
\delta &=&  2- \nu d  \\
\nu &=&\frac {2-D}{2} \ .\nn
\eea
Perturbation theory is performed about the point ($\delta=0$,
$\E=0$), i.e. ($D=1$, $d=4$). 
As the model is constructed such that the 
static limit is correctly reproduced, $Z$ and $Z_b$ are 
the renormalization-factors of the static theory \cite{ZINN}. 

Then, $\tilde Z$ is determined in order to render 
$\dot r $ finite. 

The composition of $\tilde r$ with any other operator with the 
same time argument 
is always free of  divergences, as the contraction
of $\tilde r$ with any functional of $r$ and $\tilde r$ vanishes. 
We therefore conclude that the first term in the action,
$\tilde r \dot r$ is correctly renormalized. 

Then,
$Z_\lambda$ and $Z_\eta$ are chosen to render 
the operator $\cal D$ finite.
By the same arguments as above, the last term in the action,
$\tilde r \cdot {\cal D} \tilde r$ is renormalized. 

We still have to show that also the composite operator
${\cal D}\cdot \frac{\delta H}{\delta r}$ is finite. This is a consequence
of the equation of motion obtained through variation of $J$ by 
$\tilde r(x,t)$:
\be \label{eq of motion}
 \dot r(x,t) + \left({\cal D} \cdot \frac{\delta H}{\delta r}\right)(x,t)
-2 \left( \tilde r\cdot {\cal D}\right)(x,t) =0 \ .
\ee
This relation is valid as an operator-identity.
We have already renormalized $\dot r$, ${\cal D}$, $\frac{\delta H}{\delta r}$
and $\tilde r$. Equation 
\eq{eq of motion} thus states that also the composite operator
${\cal D}\cdot \frac{\delta H}{\delta r}$ is finite. 
We can now conclude that all the terms
in the action are finite.

There are three nontrivial relations which considerably simplify
renormalization and which we are going to study now.
 First of all, due to the fact that in any interaction
vertex the field $r$ appears either as difference  $(r(x,t)-r(y,t))$ 
or as spatial derivative,  no divergence proportional to $\tilde r \dot r$
appears. (For a detailed discussion see \cite{Wiese97a}.) This means that
\be
 \tilde Z Z =1 \ .
\ee
In addition, 
there is no proper renormalization of the hydrodynamic interaction. 
Let us explain this point. Denote by 
\bea
\DynG &=& 
\int_k \left( \frac{\delta_{\alpha\beta}}{k^2}-\frac{k_\alpha k_\beta}{k^4 }\right) \rme^{i k (r(x,t) - r(y,t))} \times \nn \\
&& \times f_\alpha(x,t) g_\beta(y,t)
\eea
any hydrodynamic interaction vertex. (The dotted line represents
any polynomial in $r$ and $\tilde r$ or their derivatives.) Then 
singular configurations which give rise to a renormalization 
of the hydrodynamic interaction are those for which two interaction
vertices are contracted to one single vertex. We claim that their 
multilocal operator expansion (MOPE) does not contain a 
contribution proportional to the hydrodynamic interaction vertex:
\be
\Big( \DynH \Big| \DynG \Big) =0 
\ee
(The round dotted lines indicate points which are contracted.)
This property is due to the analytic behavior of the long range 
(hydrodynamic) interaction.
Dropping indices, the structure of such a contraction is
\bea
  \DynH  &=& \int_k \int_p k^{-2} p^{-2} \rme^{i (k+p) (r(x,t)-r(y,t))}
\times \nn \\
&& \qquad \times \rme^{kp (C(\delta x,t)+C(\delta y,t) )} 
\nn\\
&& \qquad 
+ \mbox{subdominant}
\eea
In order to obtain a long-range term a pole at $k+p=0$ is 
necessary. For $d>2$ however, the expression is analytic.
No long-range term is generated. This is easily generalized to
any order in perturbation theory.

We now use the fluctuation-dissipation theorem 
\be
\Theta(t-t') \left< r_\alpha(x,t) \dot r_\beta(y,t') \right> = \left< r_\alpha(x,t)  \left(\tilde r \cdot {\cal D}\right)(y,t') \right> \ ,
\ee
which is derived along the same lines as in \cite{Janssen92} and 
which we write down in Ito-discretization. (For other discretizations,
additional terms appear on the r.h.s.\ which cancel the contraction
of $\tilde r$ and $r$ with the same argument.)
It is valid for bare and renormalized quantities.
Inserting the definition of ${\cal D}_{\alpha \beta}$ we conclude that
\be
Z_\eta=1 \ .
\ee
In the parameterization given above, 
$Z_\lambda$ is 
\be
Z_\lambda =1+{\cal O}(\eta_R) 
\ee
as it has to vanish for $\eta_R=0$.
These relations are sufficient to completely solve for the anomalous 
exponents. We are interested in the IR-behavior. Suppose
that the coupling related to the selfavoidance has flown to its IR-fixed 
point $b_R=b^*$, what implies that also the scaling exponent $\nu(b_R)$,
defined by
\be
\nu (b_R) =
\frac{2-D}2 - \half \mu \frac{\partial}{\partial \mu}\lts_0 \ln Z
\ee
has flown to its IR-fixed point $\nu^*$.
The $\beta$-function associated to the coupling $\eta$ 
is
\bea
\beta_\eta&=& \mu\frac{\partial}{\partial \mu}\lts_0 \eta_R \nn\\
&=&\eta_R \left( -\delta +(1-d/2) \mu \frac\partial {\partial \mu}\lts_0 \ln Z + 
\mu\frac \partial {\partial \mu}\lts_0 \ln Z_\lambda \right)
\eea
Suppose now that $\eta$ has a nontrivial fixed point
$\eta^*$  for $\eta_R>0$,
i.e.\ $\beta_\eta(\eta^*,b^*)=0$ and $\frac \partial {\partial \eta_R}
\beta_{\eta}(\eta_R ,b^*)|_{\eta_R=\eta^*} >0$.
(We will show below that at leading order such a fixed point exists.)
We now express $\mu \frac\partial {\partial \mu}\sts_0 \ln Z$ 
by its value at the IR-fixed point $b_R=b^*$
\be
2(\nu^*-\nu)= -\mu \frac\partial {\partial \mu}\lts_0 \ln Z \ .
\ee
We can then solve for $\mu \frac\partial {\partial \mu}\sts_0 \ln Z_\lambda$
\be
\mu \frac\partial {\partial \mu}\lts_0 \ln Z_\lambda = \delta + (2-d) ( \nu^*-\nu)
\ee
The exponent $z$ is given by
\be
z=(D +2\nu^*- \mu \frac\partial {\partial \mu}\lts_0 \ln Z_\lambda)/{\nu^*}
\ee
The last two equations can be combined to give
\be
z=d \ .
\ee
This relation is valid as long as $\eta_R$ has flown to a non-trivial 
fixed point $\eta^*>0$. 
We will first study
the stability of the fixed point $\eta_R=0$, before 
analyzing potential fixed points for $\eta_R>0$. Without selfavoidance
this is simply the line with $\delta=0$, see figure 1.
Selfavoidance however modifies this phase-separation line.
To see this, look at $\beta_\eta$ at $\eta_R\approx 0$:
\be
\beta_\eta= \left( -\delta + (d-2) (\nu^*-\nu) \right) \eta_R + \eta_R^2 \ ,
\ee
where we used the fact that $\mu \frac\partial {\partial \mu} \ln Z_\lambda
={\cal O}( \eta_R)$.  
The stability condition for the fixed point $\eta_R=0$ is therefore 
\be
\delta<(d-2) (\nu^*-\nu)  \ .
\ee
At 1-loop order, the separating line is 
\be
\delta=(d-2) \frac{\E}{8} \ .
\ee
Numerical evaluation yields the thin line separating the 
regions with $z=d$ and $z=2+D/\nu^*$. There is however
a priori no reason to trust this estimate for membranes, i.e. $\E=4$.
We know however that in any dimension the Flory-estimate 
$\nu_{\mbox{\scriptsize Flory}} = (2+D)/(2+d)$ is 
quite a good approximation for $\nu^*$ in the fractal phase, for polymers as
well as for membranes \cite{DavidWiese96a,WieseDavid96b}.
Inserting this relation we obtain for the separatrix
\be \label{star}
	d=2(D+1) \ .
\ee
(This is the fat line between the regions with $z=d$ and $z=2+D/\nu^*$.)
Let us stress that we only use the Flory-approximation to estimate
$\nu^*$, but not any of the systematically wrong assumptions which 
have to be used to derive it.

Another possibility to get \eq{star} is to demand that the value 
of $z$ is continuous on the phase separation line. The equivalence
of the results obtained by the two methods 
is a consequence of the general structure of the renormalization group. 

We also can give a rigorous bound for the phase separation line. 
As $\nu^* \le 1$, hydrodynamics is always relevant for 
\be
d<\frac 8 {4-D}
\ee


We still have to check that $\beta_\eta$ has a fixed point for $\eta_R>0$.
At 1-loop order $Z_\lambda$ is 
\be
Z_\lambda = 1 - \Big< \DynI\Big| \GN \Big>_{\delta^{-1}} \frac {\eta_R}{\delta}
+{\cal   O}(\eta_R^2)
\ee
The diagram on the r.h.s.\ is the contraction of the hydrodynamic
interaction only. Explicitly this is 
\bea
	\Big< \DynI\Big| \GN \Big>_{\mu} \delta_{\alpha\beta} &=& \int_{x<\mu^{-1}} \int_k
\left( \frac{\delta_{\alpha \beta}}{k^2} -\frac{k_\alpha k_\beta}{k^4}\right)
\rme ^{- |x|^{2-D} k^2} \nn \\
&=& \delta_{\alpha\beta} \frac{d-1}{2d (d-2)} \frac 1\delta  \mu^{-\delta} 
\eea
The residue is thus positive
\be
\Big< \DynI\Big| \GN \Big>_{\delta^{-1}} >0 \ .
\ee
This ensures the stability of the fixed point at least for small $\delta$.

In conclusion: We have shown that the dynamical field 
theory \eq{J} for polymerized
tethered membranes including hydrodynamics is renormalizable and
that the dynamical scaling exponent $z$ is given to all orders in 
perturbation theory by $d$.

\acknowledgments
\noindent 
It is a pleasure to thank F. David, H.~W. Diehl and L. Sch\"afer 
 for stimulating discussions.

\bibliography{../citation}

\bibliographystyle{../KAY}

\end{multicols}
\end{document}